\documentclass{aastex63}
\usepackage{textcomp}
\usepackage{gensymb}
\usepackage{amssymb}
\usepackage{mathtools}
\usepackage{amsmath}
\usepackage{type1cm}
\usepackage{geometry}
\usepackage{color}
\usepackage{epsfig}
\usepackage{latexsym}
\usepackage{hyperref}
\usepackage[all]{hypcap}
\usepackage{graphicx}
\usepackage{epstopdf}
\usepackage{float}
\usepackage[caption=false]{subfig}
\usepackage{lineno}
\usepackage{placeins}
\let\splitbox\undefined
\usepackage{adjustbox}
\usepackage{array}

\received{December 10, 2020}
\revised{December 10, 2020}
\accepted{\today}
\submitjournal{ApJL}

\begin{document}

\title{ Evidence for Declination Dependence of the Ultrahigh Energy Cosmic Ray 
Spectrum in the Northern Hemisphere }

\author[0000-0001-6141-4205]{R.U. Abbasi}
\affiliation{Department of Physics, Loyola University Chicago, Chicago, Illinois, USA}

\author[0000-0001-5206-4223]{T. Abu-Zayyad}
\affiliation{Department of Physics, Loyola University Chicago, Chicago, Illinois, USA}
\affiliation{High Energy Astrophysics Institute and Department of Physics and Astronomy, University of Utah, Salt Lake City, Utah, USA}

\author{M. Allen}
\affiliation{High Energy Astrophysics Institute and Department of Physics and Astronomy, University of Utah, Salt Lake City, Utah, USA}

\author{Y. Arai}
\affiliation{Graduate School of Science, Osaka City University, Osaka, Osaka, Japan}

\author{R. Arimura}
\affiliation{Graduate School of Science, Osaka City University, Osaka, Osaka, Japan}

\author{E. Barcikowski}
\affiliation{High Energy Astrophysics Institute and Department of Physics and Astronomy, University of Utah, Salt Lake City, Utah, USA}

\author{J.W. Belz}
\affiliation{High Energy Astrophysics Institute and Department of Physics and Astronomy, University of Utah, Salt Lake City, Utah, USA}

\author{D.R. Bergman}
\affiliation{High Energy Astrophysics Institute and Department of Physics and Astronomy, University of Utah, Salt Lake City, Utah, USA}

\author{S.A. Blake}
\affiliation{High Energy Astrophysics Institute and Department of Physics and Astronomy, University of Utah, Salt Lake City, Utah, USA}

\author{I. Buckland}
\affiliation{High Energy Astrophysics Institute and Department of Physics and Astronomy, University of Utah, Salt Lake City, Utah, USA}

% ...

\begin{abstract}
Telescope Array (TA) is the largest experiment in the Northern Hemisphere studying ultrahigh energy cosmic rays. TA measurements of the cosmic ray spectrum using the surface detector have the best statistical power in the experiment, and observe the ankle of the spectrum and the high energy cutoff.  When the data are divided into two declination bands, above and below 24.8 degrees, the cutoff appears at $10^{19.64 \pm 0.04}$ ($10^{19.84 \pm 0.02}$) eV in the lower (higher) band, an energy difference of 58\%. The global significance of the difference is 4.3 standard deviations.  The lack of an instrumental cause of this difference implies it is astrophysical in nature.
\end{abstract}


\section{Introduction}
\label{section:introduction}

The Telescope Array experiment, described previously by \citet{ta:sd,ta:brlrfd,ta:mdfd}, has the aim of
studying ultrahigh energy cosmic rays.  TA measurements have been made
of the spectrum \citep{ta:sdspec_5yr} and composition
\citep{ta_comp:2018} of cosmic rays.  Searches for anisotropy
\citep{ta:hotspot_5year} and other results \citep{ta:photon_search}
have also been published.  In this paper we report on a study of the
energy of the high energy cutoff seen in different parts of the sky.

The spectrum may cut off at high energies, $E_{c} > 5 \times 10 ^{19}$ eV, as the result of
pion photoproduction in proton interactions with photons of the cosmic microwave background (CMB)
radiation,  \citep{greisen,zk}, of photodissociation of cosmic ray nuclei, or simply at the
maximal energy to which astrophysical sources can accelerate cosmic rays.

Because of a less powerful energy loss mechanism, e$^{+}$e$^{-}$
pair production in the interactions of proton with the CMB, the energy of the cutoff could also vary according to the distance to the closest sources
strong enough to produce particles at the highest energies.  The first
observation of the cutoff, at the energy predicted in 
\citet{Berezinsky:1988wi}, was made in the Northern Hemisphere sky by the High 
Resolution Fly’s Eye (HiRes) experiment, \citep{hires:spectrum}.  Two 
subsequent measurements were made by the Pierre Auger Observatory (Auger)~\citep{auger:early_spectrum} 
(at a lower energy than that predicted by \citep{Berezinsky:1988wi}, 
and in the Southern Hemisphere) and by TA~\citep{ta:sdspec_5yr} 
(at the same energy as HiRes, and also in the Northern Hemisphere).

\begin{figure*}[!ht]
	\centering
	\subfloat[]{\includegraphics[width=0.45\textwidth]{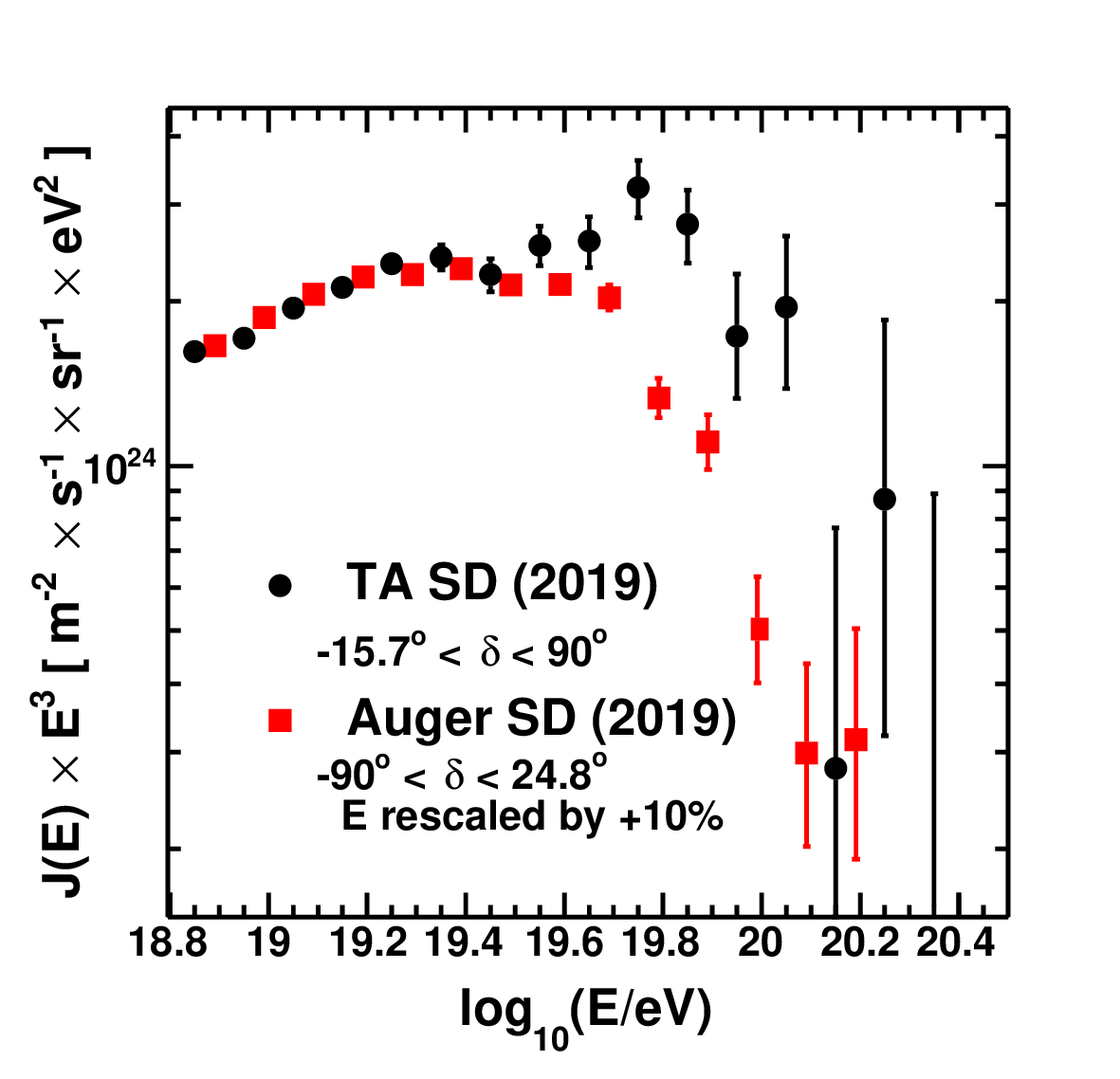} \label{figure:ta_auger_fullsky}}
	\subfloat[]{\includegraphics[width=0.45\textwidth]{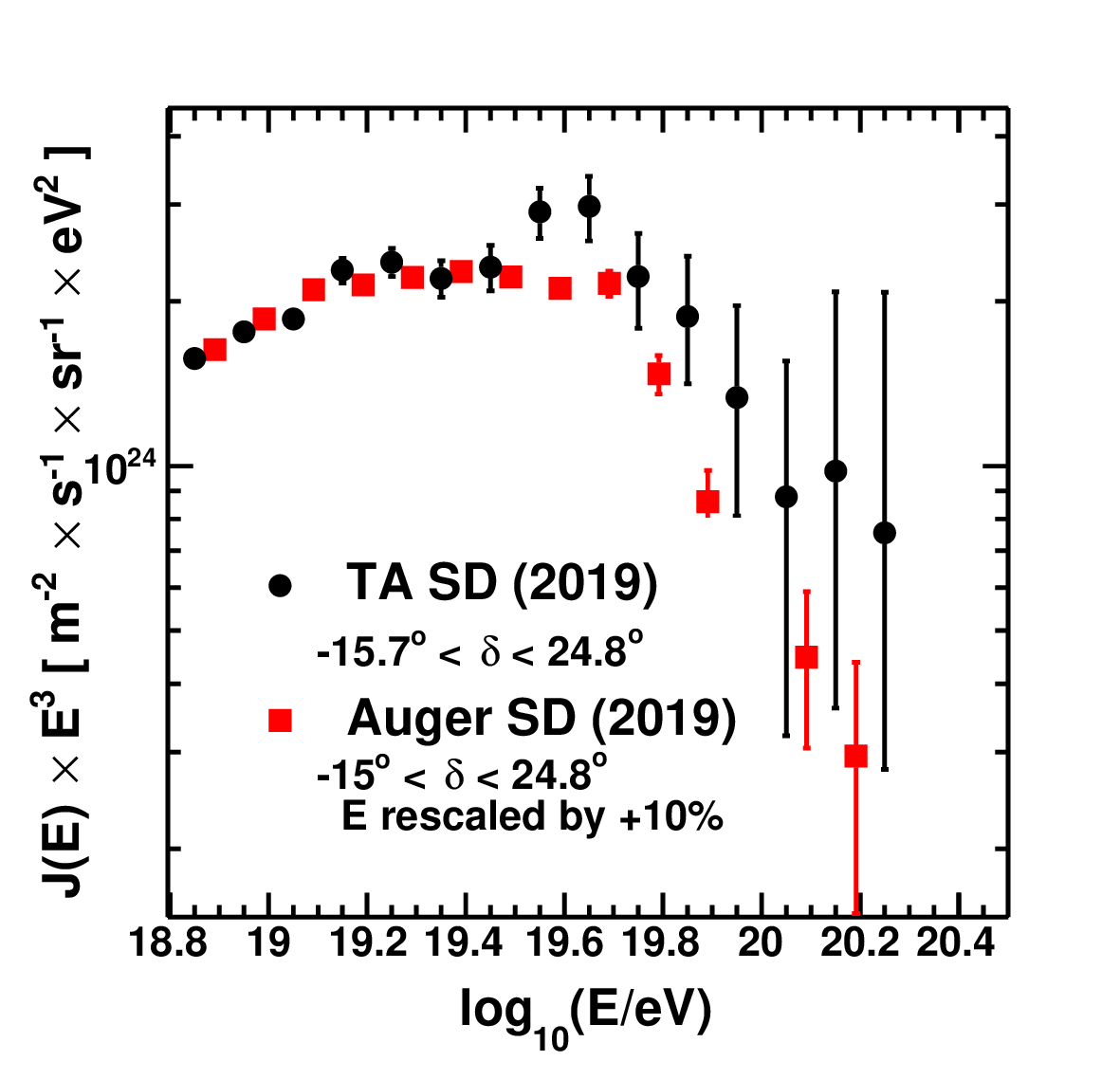} \label{figure:ta_auger_common}}
	\caption{ \protect\subref{figure:ta_auger_fullsky} TA and Auger full sky 
	spectra as of 2019, with the Auger energy scale increased by $10\%$.
	\protect\subref{figure:ta_auger_common} TA and Auger surface detector 
	spectra measured in the common declination band. }
	\label{figure:spectra}
\end{figure*}

This study followed originally from a working group \citep{uhecr2016_specwg} 
formed from members of the TA and Auger collaborations to consider differences in 
the two experiments’ spectrum measurements.  Auger has a surface detector (SD) of about 1600 Cherenkov tanks located on the Pampa Amarilla in Argentina.  Part of the 
spectral difference seems to come from an energy scale difference between the 
two experiments, but even if a correction is made for this, the cutoff appears 
at different energies in the two experiments’ results.  
Figure~\ref{figure:ta_auger_fullsky} shows the two spectra after making the 
correction for energy scale difference.  To isolate effects that could produce the 
difference in cutoff energy, the working group then studied the spectra 
measured in the declination band common to both experiments.  This band 
stretches from $-15$ degrees of declination (the lowest observable by TA) to $24.8$ 
degrees (the highest observable by Auger).  Figure~\ref{figure:ta_auger_common} 
shows the two spectra in the common declination band, again after correction for
energy scale difference.  The two measurements of the energy of the
cutoff are consistent in the common declination band.  These surprising results indicate that the spectrum in the Northern Hemisphere varies with declination.

In Section ~\ref{section:TASD} we describe the surface detector of TA.  In 
Section ~\ref{section:TA_Auger_CommonBand} we examine further the TA and Auger 
spectra in the common declination band. In Section ~\ref{section:results} we 
present the results of the study.  In Section ~\ref{section:systematics_search} 
we describe a search for an instrumental cause of the cutoff energy difference. 
And in Section ~\ref{section:summary} we present a summary.

\section{The Surface Detector of the Telescope Array}
\label{section:TASD} 

Each of the 507 detectors of the SD consists of two scintillation
counters, each of 3 m$^{2}$ area, viewed independently by photomultiplier
tubes.  Scintillation light is collected and brought to the phototubes
by wavelength-shifting fibers embedded in grooves on the scintillator.
Each channel undergoes a flash analog-to-digital (FADC) conversion
every 20 ns, and the result is saved for several seconds.  A radio
system connects each counter with one of three radio control towers
spaced around the array, each controlling about 1/3 of the array.
Once per second the tower polls each counter in its area to read out
the times that the counter has recorded a signal of strength 3 times
what a minimum ionizing particle (MIP) would provide \citep{ta:sd}.  If three
adjacent counters report signals within 8 $\mathrm{\mu}$s of each other the
array is triggered.  Then the radio tower commands each counter that
has a signal of strength 1/3 MIP or more to provide its FADC traces.
These are written into the data event record.  The radio towers
communicate with each other when triggered counters are near the area
boundaries to collect all possible detector signals.

Every 10 minutes each detector is commanded to upload a variety of
information about its performance.  In 10-minute periods each
scintillation counter accumulates a histogram of pulse heights for
single muon events by triggering on coincidences between the two
counters.  This histogram forms the basis for calibrating the counter
signals in MIP units.  The peak height and RMS value are used in the
data analysis and Monte Carlo simulation.

\begin{figure*}[!ht]
	\centering
	\subfloat[]{\includegraphics[width=0.45\textwidth]{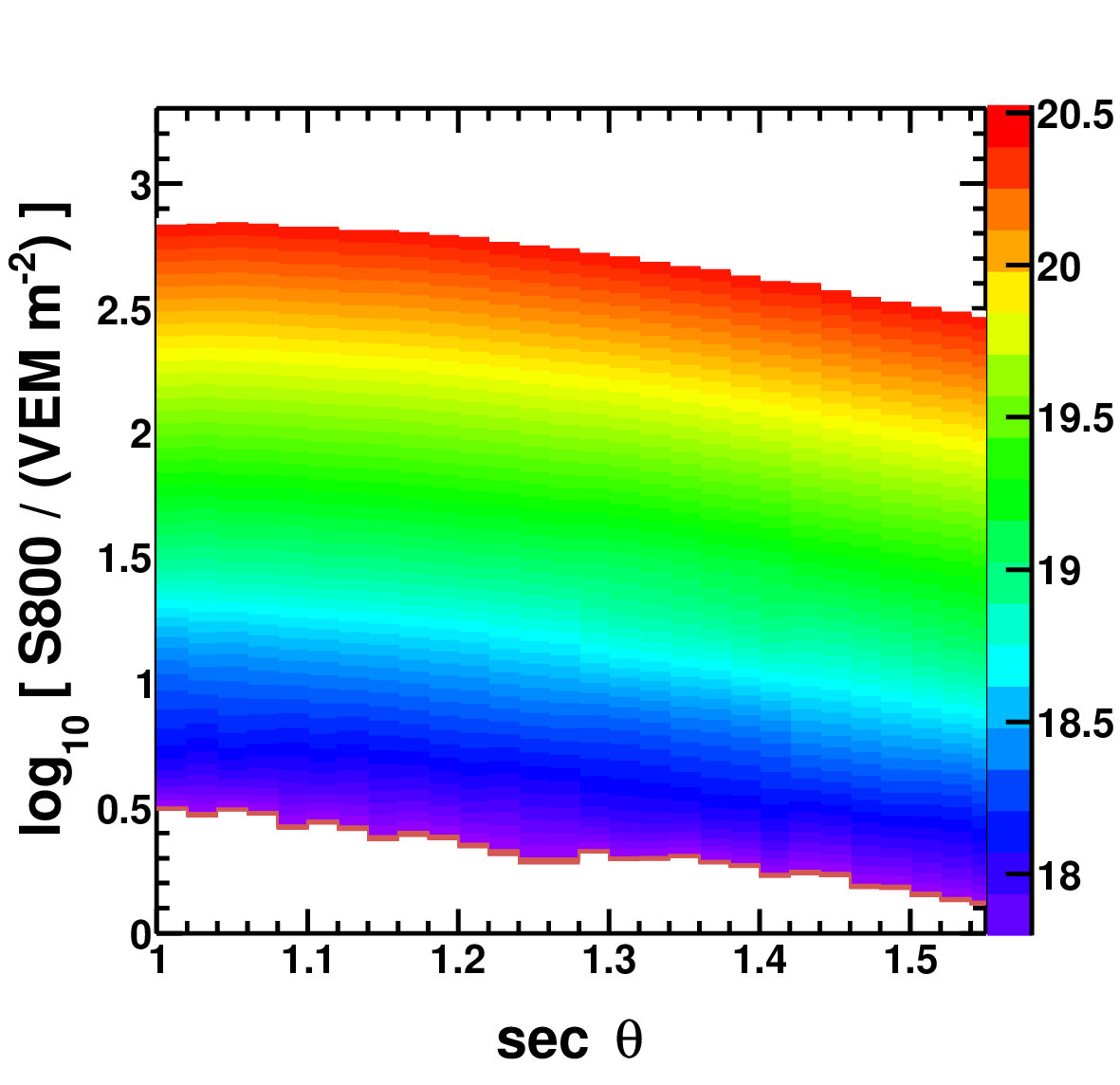}
		\label{figure:sdentable}}
	\qquad
	\subfloat[]{\includegraphics[width=0.45\textwidth]{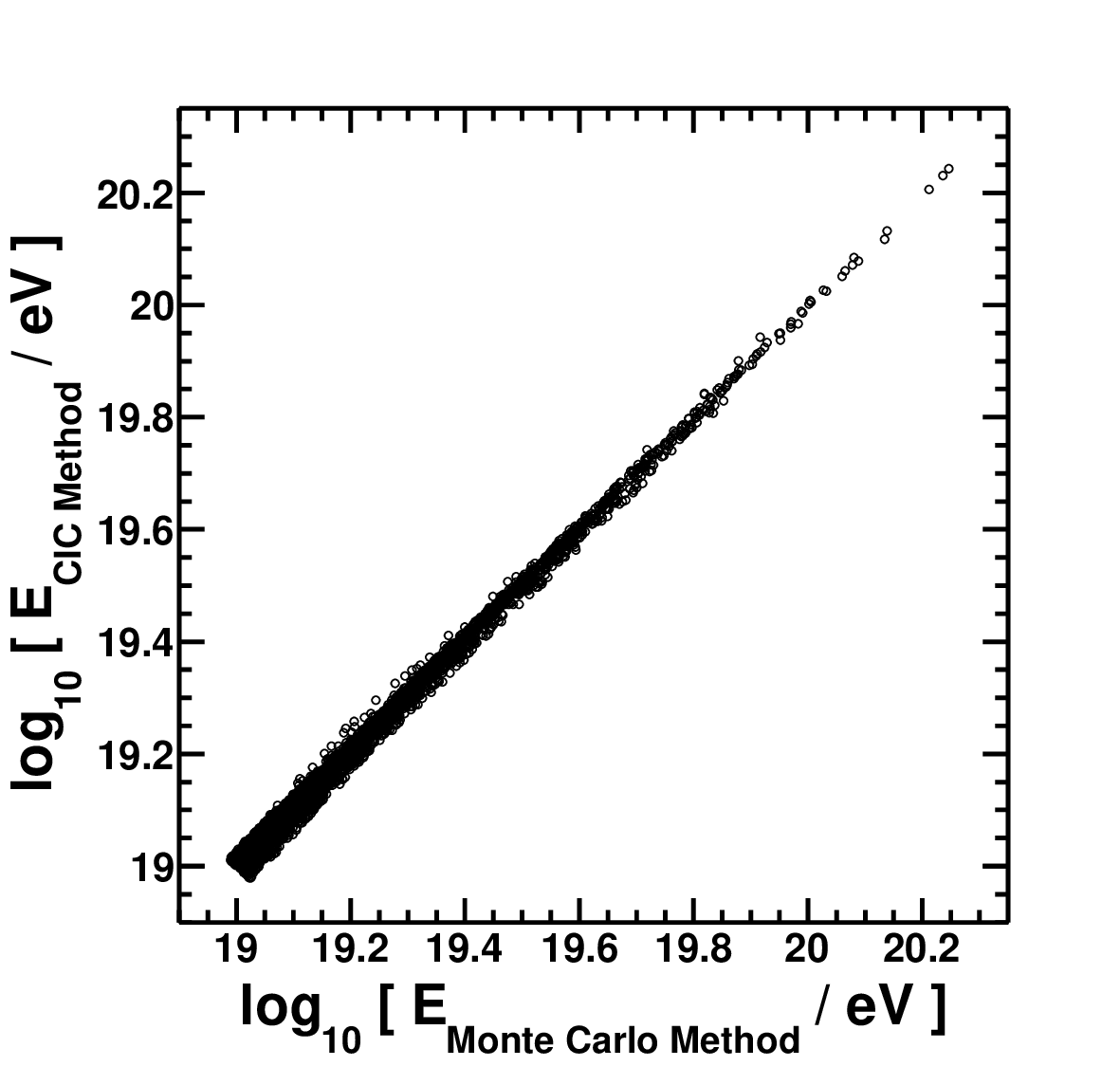}
		\label{figure:cic_vs_mc_scatter}}
	\caption{ TA SD event energies were reconstructed using an
	energy estimation table \citep{icrc2015_spectrum_summary} derived from the Monte Carlo, shown in 
	\protect\subref{figure:sdentable}, and using the Constant Intensity Cut 
	method \citep{method:CIC}. The scatter plot in 
	\protect\subref{figure:cic_vs_mc_scatter} shows the 
	comparison between the two different reconstruction methods applied to the 
	TA SD data.  The standard deviation of the logarithmic difference of the 
	energies reconstructed by the two methods was $\sim$ 3\%. }
	\label{figure:tasd_energy}
\end{figure*}

The data analysis starts with identifying clusters of counters struck
by the particles in a cosmic ray air shower.  The times when counters
are struck are used in reconstructing the arrival direction of the cosmic ray.
The signal size and times are used to calculate the center of the
shower, and counters’ pulse heights and distances from the center are
used to calculate $\mathrm{S}_{800}$, the signal at a distance of 800 m
from the shower center, by interpolation.  To reconstruct the energy
of the cosmic ray primary particle, the $\mathrm{S}_{800}$ value and zenith angle are compared to
Monte Carlo simulations of the SD \citep{ta:sdspec_5yr}.  Figure~\ref{figure:sdentable} shows
the relation between $\mathrm{S}_{800}$ and zenith angle for Monte Carlo events of
various energies.  As a check on this method of energy reconstruction
we also use the constant-intensity-cut (CIC) reconstruction method 
\citep{method:CIC}. Figure~\ref{figure:cic_vs_mc_scatter} shows a scatter plot 
of the two energy values for each event:  CIC based vs Monte Carlo based.  The 
agreement between the two methods is excellent.  In both cases the SD energy was compared with the energy of the TA fluorescence detector (FD) for
events where a good reconstruction is possible by both detectors.
Since the FD energy determination is calorimetric this gives us a
robust energy reconstruction for SD events.  The TA SD is approximately 100\%
efficient for cosmic rays energies above about 10$^{18.9}$ eV.

\section{The TA and Auger Spectra in the Common Declination Band}
\label{section:TA_Auger_CommonBand}

In Figure~\ref{figure:ta_auger_common} we saw that the cutoff energies of the 
TA and Auger spectra are consistent in the common declination band.  The
two spectra are not identical, however, and in this section we examine
the origin of the remaining difference.

\begin{figure*}[!ht]
	\centering
	\subfloat[]{\includegraphics[width=0.45\textwidth]{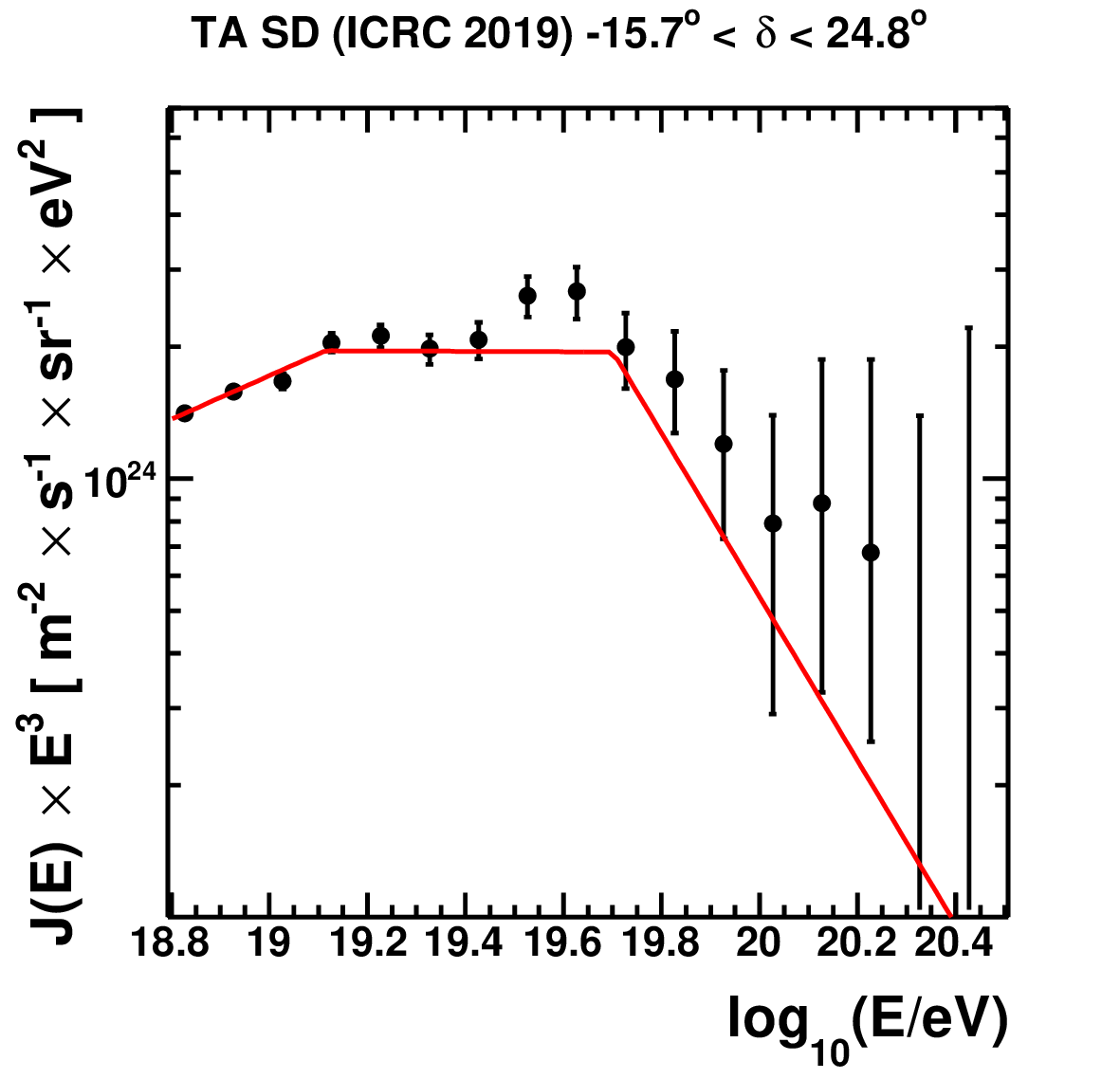}
		\label{figure:fit_ta}}
	\subfloat[]{\includegraphics[width=0.45\textwidth]{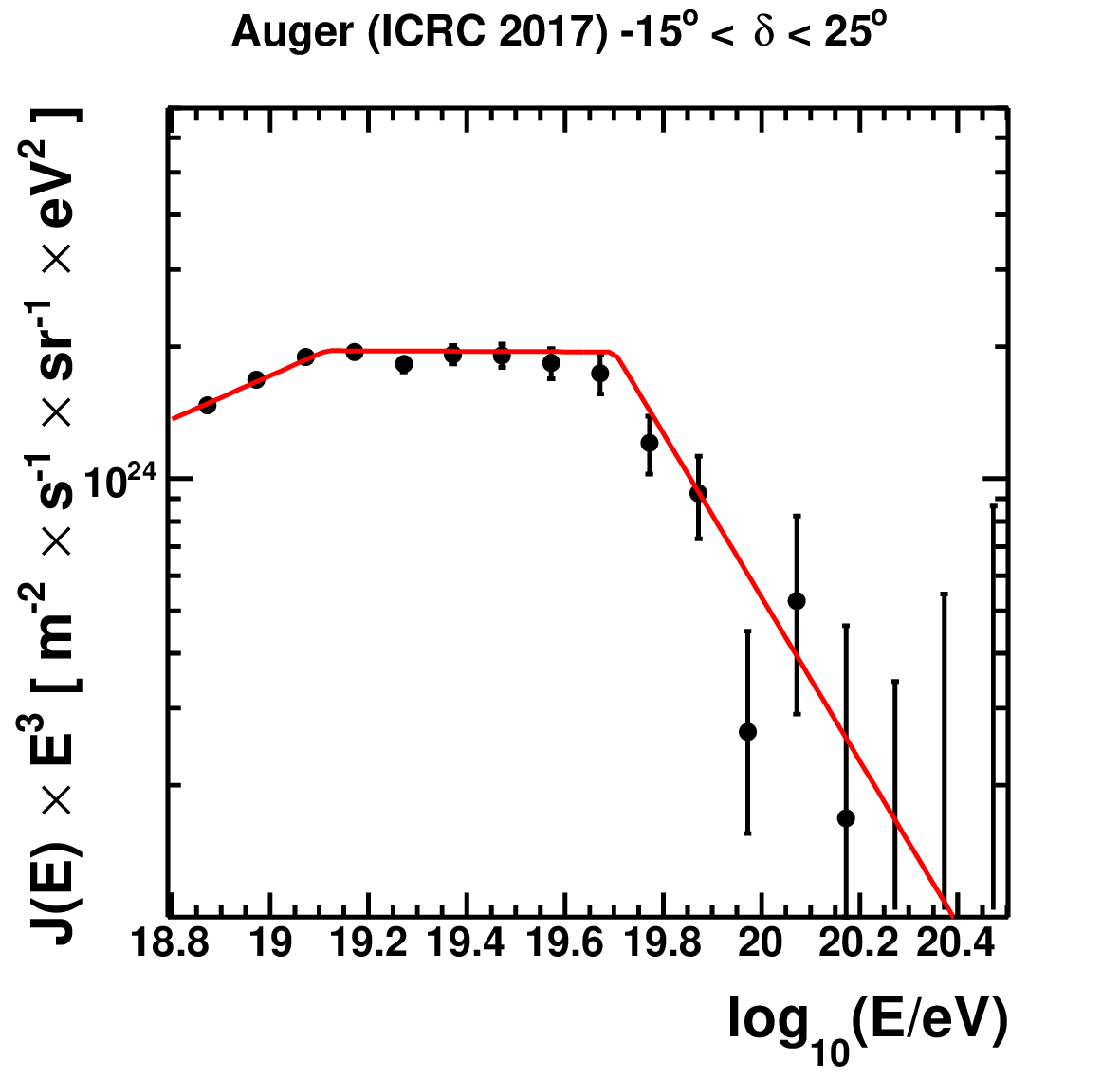}
		\label{figure:fit_auger}}
	\caption{ Fit to the TA and Auger spectra measured in the common 
	declination band of the same broken power law function (with breaks found 
	at $\sim$ $10^{19.1}$ and $\sim$ $10^{19.7}$ eV)  with correction for a 
	constant ($+5.2\%$ for Auger, $-5.2\%$ for TA) overall energy scale shift.  The log likelihood sum over nonzero event bins for this fit is 38.5 over 25 degrees of freedom.
		\protect\subref{figure:fit_ta} TA spectrum points on top of the joint 
		fit function.
		\protect\subref{figure:fit_auger} Auger spectrum points on top of the 
		joint fit function.  }
	\label{figure:fit_ta_auger}
\end{figure*}

In order to characterize the difference between the Auger and TA spectra
in the common declination band, we performed a simultaneous fit to the
two spectra.  This fit assumes there are three power law sections
separated by two break points (we include the recently seen shoulder
at $10^{19.2}$ eV \citep{auger:spectrum_features_2020} in our fits).  The result 
is shown in Figure~\ref{figure:fit_ta_auger}.  It has a chi-squared of 38.5 
for 25 degrees of freedom.  Because the Auger spectrum 
is based on about six times the number of events as TA, the fit follows the Auger 
spectrum more closely than that of TA.  Above an energy of 10$^{19.5}$ eV the TA
data points can be seen to be about 15\% higher than those of Auger.

To understand how often such a difference could arise randomly, we
constructed a model of the TA and Auger spectrum measurements.  It
included the spectrum as found in the simultaneous fit from the
previous paragraph, the energy resolution, the correction for bin
migration, the anisotropy dipole seen by the Auger collaboration, and
the excess of events called the {\it hotspot} seen by the TA collaboration.
We note that the TA spectrum in the common declination band has highest 
statistical power at the northern end of the band where the hotspot enters, 
while the Auger spectrum is weighted at the southern end away from the
hotspot.  We threw the model with the observed number of TA and Auger
events $100\,000$ times and found that $\sim$10\% of the time there was a
divergence between the two spectra of similar size or larger as that of the
data.  We conclude that the observed difference between the TA and Auger
spectra in the common declination band may be due to a statistical
fluctuation.

One could imagine an energy dependent change in the energy resolution producing a resolution function with a longer tail than Auger's above $10^{19.4}$~eV. However, a straightforward application of such a hypothesis to the data with $\delta > 24.8\degree$ would lead to a contradiction, since that spectrum has a steeper fall off than the $\delta < 24.8 \degree$ data. The remaining possibility that the resolution function has both an energy and a declination dependence (via the zenith angle) is ruled out in Section~\ref{section:systematics_search}.

\section{Results: The Cutoff Energy Varies with Declination}
\label{section:results}

\begin{figure*}[!ht]
	\centering
    \begin{tabular}{cc}
	\subfloat[]{\includegraphics[width=0.61\textwidth]{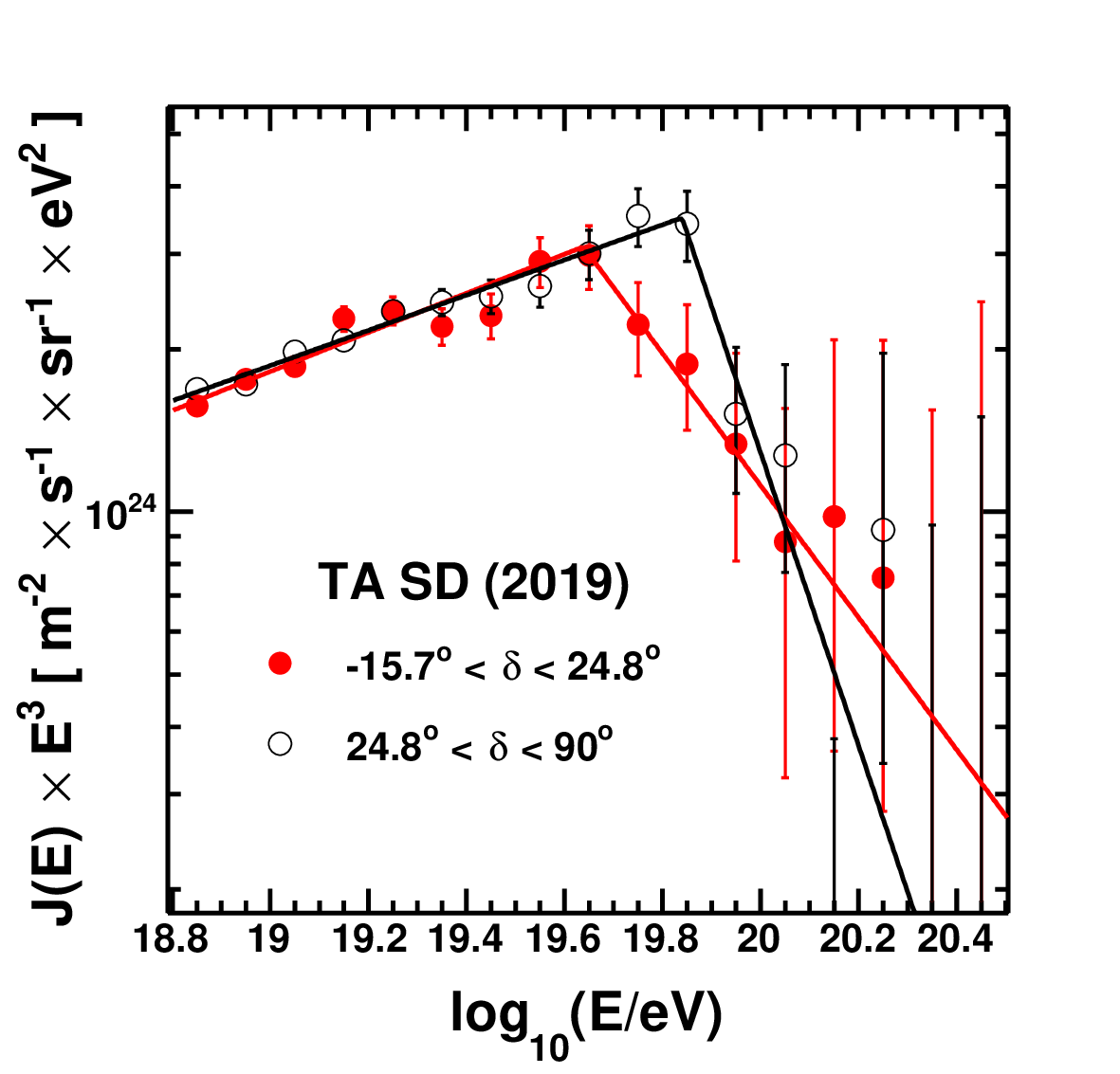} \label{figure:tasd_bands_11years}}
	&
	\adjustbox{valign=b}{\begin{tabular}{@{}c@{}}
	\adjustbox{valign=b}{\subfloat[]{\includegraphics[width=0.23\textwidth]{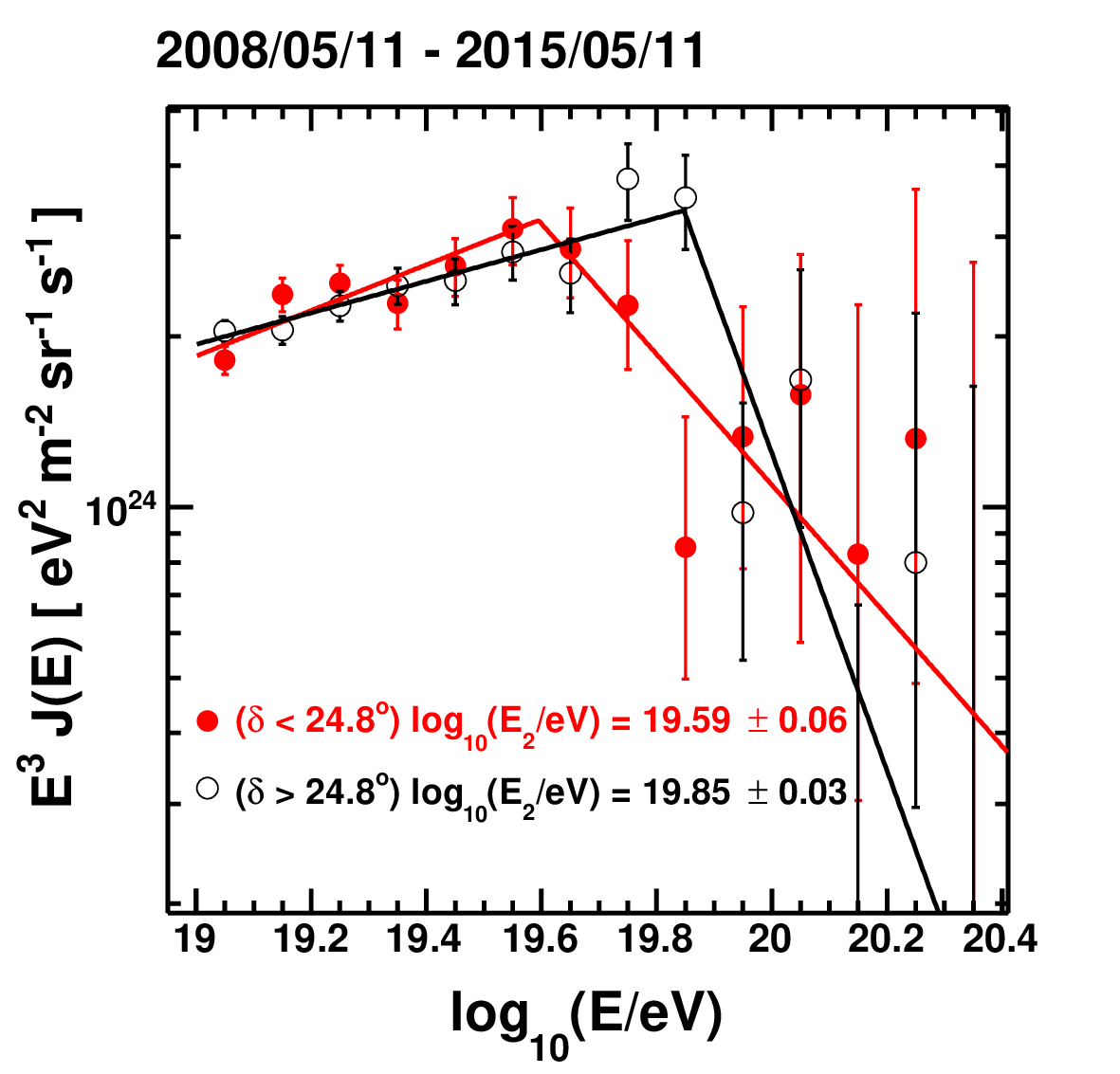} \label{figure:tasd_bands_7years}}} \\
	\subfloat[]{\includegraphics[width=0.23\textwidth]{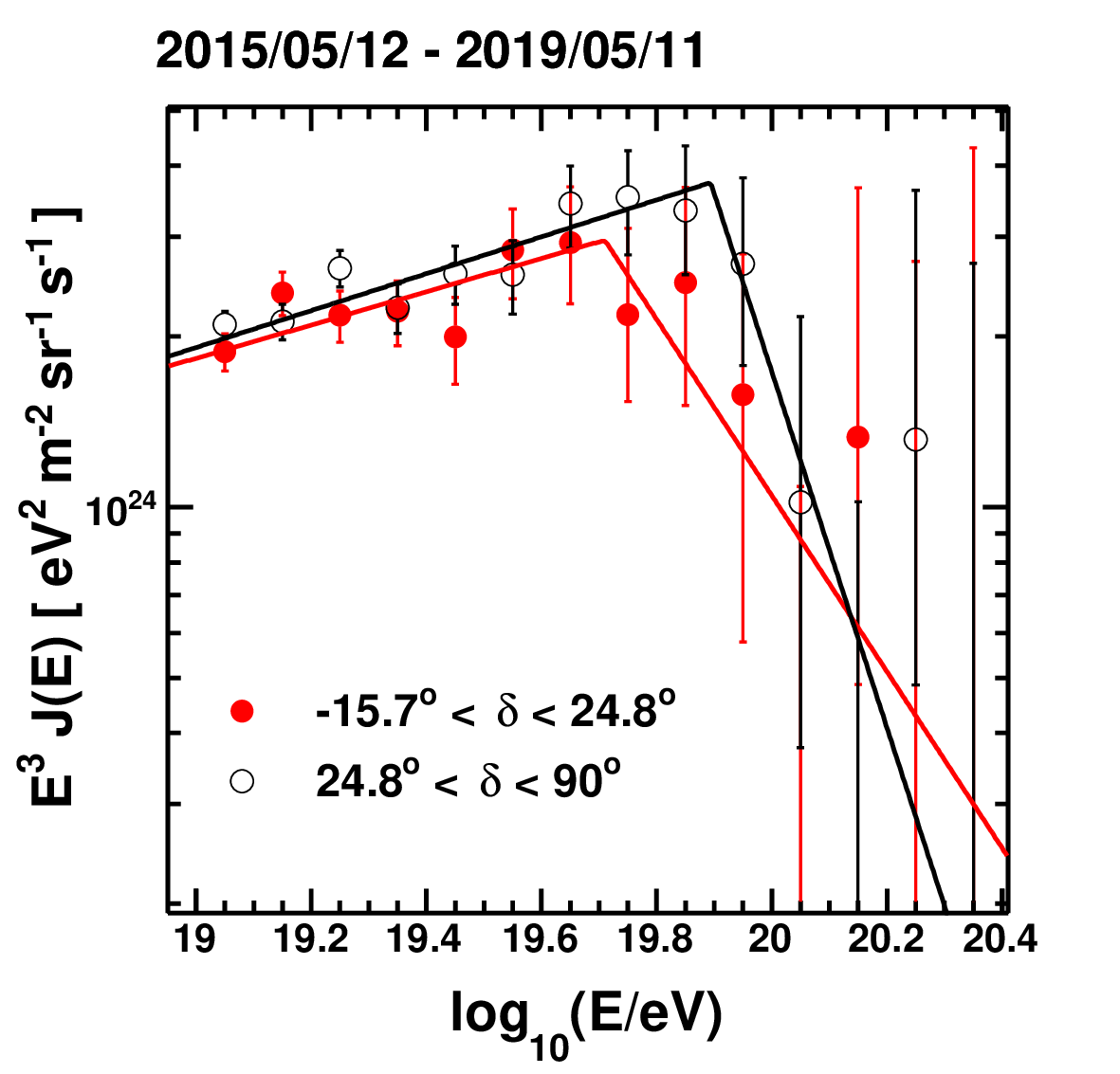} \label{figure:tasd_bands_4years}} 
	\end{tabular}}
	\end{tabular}
	\caption{ TA SD spectra measured in the two declination bands, 
	$-15.7\degree < \delta <
		24.8\degree$ and $24.8\degree < \delta < 90\degree$,
		\protect\subref{figure:tasd_bands_11years} for the entire available data set (2008/05/11--2019/05/11)
		\protect\subref{figure:tasd_bands_7years} for the first 7 years of TA SD 
		data (2008/05/11--2015/05/11) and \protect\subref{figure:tasd_bands_4years} for the remaining 
		4 years of data (2016/05/11--2019/05/11).  The declination dependence effect is qualitatively 
		seen in all three data sets.  In \protect\subref{figure:tasd_bands_11years}, for $-15.7\degree < \delta <
		24.8\degree$, the spectral indices are $-2.64 \pm 0.04$ and $-4.2 \pm 0.3$ before and after the break point at $10^{19.64
      \pm 0.04}$ eV.  For $24.8\degree < \delta < 90\degree$, the spectral indices are $-2.67 \pm 0.03$ and $-5.71
    \pm 0.6$, respectively, and the break point is at $10^{19.84 \pm 0.02}$ eV.  For \protect\subref{figure:tasd_bands_7years} and \protect\subref{figure:tasd_bands_4years} the numbers are similar and can be found in \citet{Ivanov:2019qiu,icrc2019_taspec}. }
	\label{figure:tasd_7_and_4_years}
\end{figure*}

Figure~\ref{figure:tasd_bands_7years} shows the spectra of ultrahigh energy 
cosmic rays measured with the SD of the Telescope Array above and below 24.8 
degrees in declination.  Data collected during the first 7 years of TA operation are included in this figure, corresponding to the time the TA collaboration first noted this effect \citep{icrc2017_specwg}.  In the figure a broken power law fit was performed to the data to quantify the energy of the cutoff.  The characteristics shown in Figure~\ref{figure:tasd_bands_11years} are (1) the cutoff energy in the lower declination band, shown in red, agrees within 1/2 standard deviation with the Auger result; (2) the cutoff energy in the higher declination band is higher than that seen by TA in the whole sky; and (3) the data in both declination bands are well fitted by a broken power law model.

We calculate the significance of the difference in cutoff energies first by using the
fit values of the cutoff energies and uncertainties \citep{icrc2017_specwg} and get a
significance of 3.9 standard deviations.  This is called the “local significance.”  To determine the probability that this difference could arise by chance, a Monte Carlo calculation was carried out. In this calculation trials with the same number of events as the data
were generated, with the events placed randomly according to the
TA aperture, and with energies chosen according to the whole-sky
spectrum of TA.  The spectrum was then calculated for the two regions
of declination, and trials were counted in which the three characteristics of the
spectra listed in the previous paragraph were true and the cutoff
energy was different in the two regions as much as or more than the
data.  The result is called the global significance and
is 3.5 standard deviations \citep{Ivanov:2019qiu}.

Figure~\ref{figure:tasd_bands_4years} shows the spectra in the two declination 
intervals for years 8--11 of TA operation.  Here we see that the effect 
persists.  Figure~\ref{figure:tasd_bands_11years} shows the spectra for the 
whole 11 years of TA data \citep{icrc2019_taspec}. The local significance is 4.7 standard deviations, and the global significance is 4.3 standard deviations.  In 11 years of TA SD data, there are $10\,365$ events above $10^{18.8}$~eV, and the analysis details are described in \citet{icrc2019_taspec}.

\section{Search for an Instrumental Cause of the Effect}
\label{section:systematics_search} 

In the previous section we saw that there is strong evidence that the
spectrum of ultrahigh energy cosmic rays, measured by the TA surface
detector, varies with declination.  The question now arises, is this
an instrumental effect arising from some feature of the TA SD analysis
or is the effect astrophysical in origin.

We have searched for an instrumental cause in many ways such as a
possible nonlinearity in FD event reconstruction or a nonlinearity in
FD/SD energy comparison and found no evidence for these effects \citep{uhecr2018_specwg}.

\begin{figure*}[!ht]
	\centering
	\subfloat[]{\includegraphics[width=0.45\textwidth]{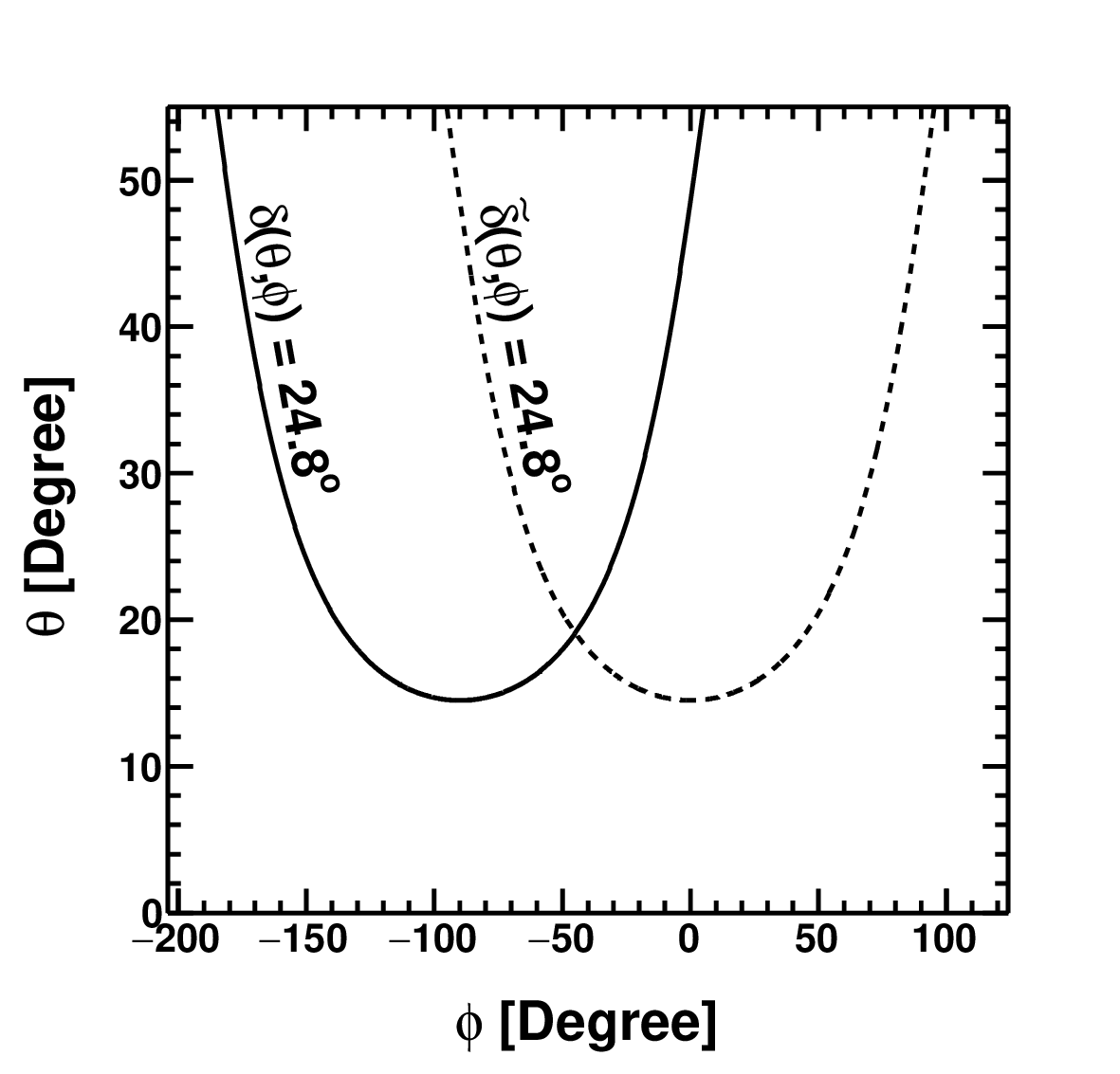}
		\label{figure:ushape_explain}}
	\hfill
	\subfloat[]{\includegraphics[width=0.45\textwidth]{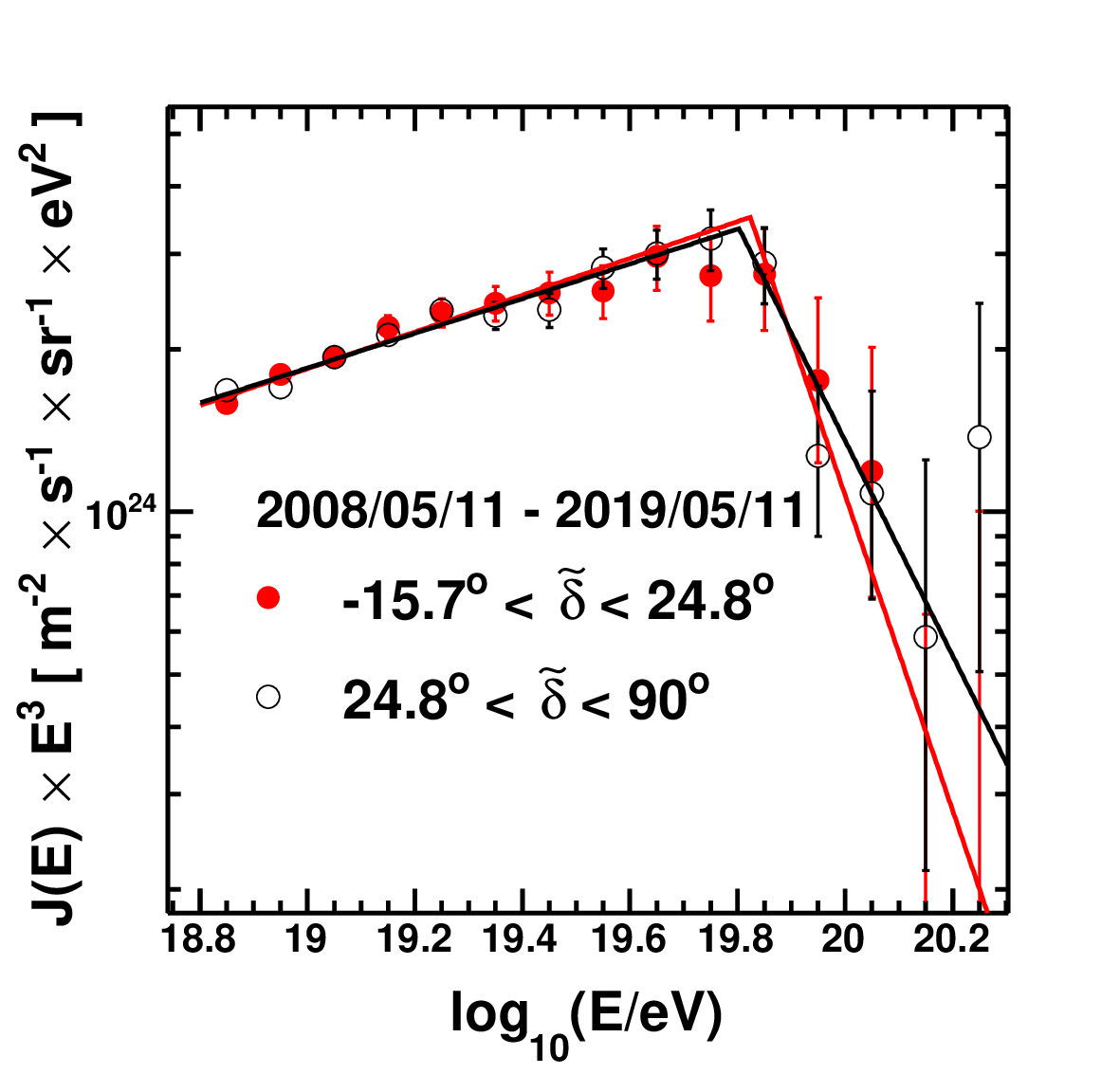}
		\label{figure:ushape_check}}
	\caption{ \protect\subref{figure:ushape_explain} Angles $\theta$ and $\phi$ 
	represent the zenith and azimuthal angles of the event arrival direction in 
	the local horizontal coordinates of the TA SD.  A contour of constant 
	declination $\delta(\theta,\phi)=24.8^\degree$ is represented by the solid 
	curve, while the dotted curve shows the corresponding contour of the constant 
	{\it modified declination}, $\tilde{\delta}(\theta,\phi) = 
	\delta(\theta,\phi-90^\degree)$. \protect\subref{figure:ushape_check} shows 
	the TA SD spectrum measured for $\tilde{\delta} < 24.8^\degree$ and 
	$\tilde{\delta} > 24.8^\degree$.  The two spectra agree to within $2\%$. }
	\label{figure:tasd_phase_space_explain_and_check}
\end{figure*}

A stringent test of a north-south spectrum difference is to search for
an east-west spectrum difference.  The grid of the TA SD layout lies
along north-south and east-west axes, so an instrumental effect would
appear in the east-west direction also.  In particular, an acceptance
effect, which could be a function of zenith angle, or a nonlinearity
in energy assignment would also appear in an east-west direction.
Figure~\ref{figure:ushape_explain} shows a plot of zenith angle vs.\ azimuthal 
angle with the left u-shaped curve indicating the locus of points whose 
declination is 24.8 degrees.  Events inside and above the curve are below 24.8
degrees.  A simple translation of the formula for this curve allows us
to look at a similar difference in an east-west direction, as also
shown in the figure.  The test consists of measuring the spectrum
using events above and, independently, below the curve on the right in
Figure~\ref{figure:ushape_explain}.  An instrumental effect would make the 
spectra different.  If the spectra are the same, it would strongly support an 
astrophysical origin of the north-south difference in cutoff energies, 
as the whole sky rotates through the east and west regions.  Figure 
~\ref{figure:ushape_check} shows the test result.  The two spectra have the 
same power laws and cutoff energies, and the point-to-point ratio of the 
spectra is 1.00 $\pm$ 0.02.  This would not be the case if the difference in 
cutoff energies above and below a declination of 24.8 degrees were due to
acceptance or energy scale.  We conclude that this is strong support
for the interpretation that the spectrum of ultrahigh energy cosmic
rays is different in lower and higher declination bands due to an
astrophysical effect.

\newpage
\section{Summary}
\label{section:summary}

The Telescope Array and Pierre Auger collaborations established a working group to study the differences in their measurements of the spectrum of cosmic rays \citep{uhecr2018_specwg}.  After making an energy scale adjustment of about 10\% to bring the ankle region of the two spectra into alignment, there remains a difference in the two spectra in the energy of the high-energy cutoff.  The working group then examined the two experiments' spectra in the band of declination covered by both experiments and found that there the cutoff energies agreed.  Although not anticipated by the two collaborations, this is not an impossible result since the energy of the cutoff could vary across the sky due to the maximum energy of different sources, or their distances from Earth.

This result prompted the Telescope Array collaboration to measure the
spectrum in the northern part of the sky (above the 24.8 degree limit
of the common declination band).  The cutoff energy was found to be
higher in that region.  In the 11 years of TA data that have been
collected, the global significance of the difference is 4.3 standard
deviations.

A comprehensive search for an instrumental effect to explain the
difference has failed to find one, and both the TA collaboration and
the joint TA-Auger working group studying the spectrum of ultrahigh
energy cosmic rays have concluded that the variation of the cutoff
energy with declination is an astrophysical effect.

In a recent paper, \citet{auger:spectrum_measurement}, the Auger collaboration 
reported that the spectrum in the Southern Hemisphere is independent of 
declination, which they say may indicate that the sources are common across the
Southern Hemisphere sky.  The present paper reports strong evidence that the
sky in the Northern Hemisphere is different: the spectrum varies with
declination.  This may be ascribed to sources that have a higher
maximum energy or perhaps are closer to the earth.

\acknowledgments
The Telescope Array experiment is supported by the Japan Society for
the Promotion of Science(JSPS) through
Grants-in-Aid
for Priority Area
431,
for Specially Promoted Research
JP21000002,
for Scientific  Research (S)
JP19104006,
for Specially Promoted Research
JP15H05693,
for Scientific  Research (S)
JP15H05741, for Science Research (A) JP18H03705,
for Young Scientists (A)
JPH26707011,
and for Fostering Joint International Research (B)
JP19KK0074,
by the joint research program of the Institute for Cosmic Ray Research (ICRR), The University of Tokyo;
by the Pioneering Program of RIKEN for the Evolution of Matter in the Universe (r-EMU);
by the U.S. National Science
Foundation awards PHY-1404495, PHY-1404502, PHY-1607727, PHY-1712517, PHY-1806797, PHY-2012934, and PHY-2112904;
by the National Research Foundation of Korea
(2017K1A4A3015188, 2020R1A2C1008230, \& 2020R1A2C2102800) ;
by the Ministry of Science and Higher Education of the Russian Federation under the contract 075-15-2020-778, IISN project No. 4.4501.18, and Belgian Science Policy under IUAP VII/37 (ULB). This work was partially supported by the grants ofThe joint research program of the Institute for Space-Earth Environmental Research, Nagoya University and Inter-University Research Program of the Institute for Cosmic Ray Research of University of Tokyo. The foundations of Dr. Ezekiel R. and Edna Wattis Dumke, Willard L. Eccles, and George S. and Dolores Dor\'e Eccles all helped with generous donations. The State of Utah supported the project through its Economic Development Board, and the University of Utah through the Office of the Vice President for Research. The experimental site became available through the cooperation of the Utah School and Institutional Trust Lands Administration (SITLA), U.S. Bureau of Land Management (BLM), and the U.S. Air Force. We appreciate the assistance of the State of Utah and Fillmore offices of the BLM in crafting the Plan of Development for the site.  Patrick A.~Shea assisted the collaboration with valuable advice and supported the collaboration’s efforts. The people and the officials of Millard County, Utah have been a source of steadfast and warm support for our work which we greatly appreciate. We are indebted to the Millard County Road Department for their efforts to maintain and clear the roads which get us to our sites. We gratefully acknowledge the contribution from the technical staffs of our home institutions. An allocation of computer time from the Center for High Performance Computing at the University of Utah is gratefully acknowledged.

\newpage
\FloatBarrier

\bibliographystyle{aasjournal}
\bibliography{specdec}

\end{document}